\documentclass[%
reprint,
superscriptaddress,
 amsmath,
 amssymb,
 aps,
pre,
]{revtex4-1}

\usepackage{graphicx}% Include figure files
\usepackage{dcolumn}% Align table columns on decimal point
\usepackage{bm}% bold math
\usepackage{hyperref}% add hypertext capabilities
\begin{document}

\preprint{APS/123-QED}

\title{Sub-Jamming Transition in Binary Sphere Mixtures}

\author{Ishan Prasad}

\affiliation{%
 Department of Chemical Engineering, University of Massachusetts, Amherst, Massachusetts 01003, USA
}%

\author{Christian Santangelo}
\affiliation{
 Physics Department, University of Massachusetts, Amherst, Massachusetts 01003, USA
}%

\author{Gregory Grason}
\email{grason@mail.pse.umass.edu}
\affiliation{%
Department of Polymer Science and Engineering, University of Massachusetts, Amherst, Massachusetts 01003, USA
}%

\date{\today}% It is always \today, today,

\begin{abstract}
We study the influence of particle size asymmetry on structural evolution of randomly jammed binary sphere mixtures with varying large-sphere/small-sphere composition.   Simulations of jammed packings are used to assess the transition from large-sphere dominant to small-sphere dominant mixtures.  For weakly asymmetric particle sizes, packing properties evolve smoothly, but not monotonically, with increasing small sphere composition, $f$.  Our simulations reveal that at high values of ratio $\alpha$ of large to small sphere radii, ($\alpha\geq \alpha_c \approx 5.75$) evolution of structural properties such as packing density, fraction of jammed spheres and contact statistics with $f$ exhibit features that suggest a sharp transition, either through discontinuities in structural measures or their derivatives.  We argue that this behavior is related to the singular, composition dependence of close-packing fraction predicted in infinite aspect ratio mixtures $\alpha\rightarrow\infty$ by the Furnas model, but occurring for finite values range of $\alpha$ above a critical value, $\alpha_c \approx 5.75$.  The existence of a sharp transition from small- to large-$f$ values for $\alpha\geq \alpha_c$ can be attributed to the existence of a {\it sub-jamming transition} of small spheres in the interstices of jammed large spheres along the line of compositions $f_{sub}(\alpha)$.  We argue that the critical value of finite size asymmetry $\alpha_c \simeq 5.75$ is consistent with the geometric criterion for the transmission of small sphere contacts between neighboring tetrahedrally close packed interstices of large spheres, facilitating a cooperative sub-jamming transition of small spheres confined within the disjoint volumes. 
%\begin{description}
%\end{description}
\end{abstract}

\pacs{Valid PACS appear here}% PACS, the Physics and Astronomy
\maketitle

%\tableofcontents

\section{\label{sec:intro}Introduction
%\protect\\ The line
%break was forced \lowercase{via} \textbackslash\textbackslash
}

Disordered hard sphere packings are useful as models of diverse systems, from molecular liquids~\cite{Bernal1959, Chaikin2000, Hansen2006} and glasses~\cite{Parisi2005}, to colloids~\cite{Torquato2002a,Aste2008}, amorphous metals~\cite{Clarke1988} and granular materials~\cite{Mehta1994, Torquato2010}.  In these contexts, hard sphere packings provide a framework for connecting the microstucture of a multi-particle system to emergent macroscopic physical properties (e.g. structure, thermodynamics, mechanics) through robust, often universal, principles of sphere-packing geometry.

Physical realizations of disordered sphere packings occur in random mixtures of athermal spheres, or otherwise when particle suspensions are quenched to high-density sufficiently rapidly to avoid crystallization and become trapped in mechanically-stable (i.e. ``jammed"), yet positionally-disordered, random close packed (RCP) state.   While seemingly random, these jammed amorphous packings have very robust, albeit poorly understood, properties, the subject of intense scrutiny beginning from pioneering studies of Bernal~\cite{Bernal1964}.  A range of experiments and simulations show that randomly-jammed packings of monodisperse spheres exhibit a surprisingly well-defined  volume fraction, $\Phi^{mono}_{\rm cp} \simeq 0.634-0.65$, notwithstanding their evident lack of long-range order and numerous differences in the kinetic approach to jamming~\cite{Anikeenko2007, Atkinson2013, Finney1970, Bernal1967, Clusel2009, Donev2004, Goodrich2012, Jodrey1985}.  Beyond density, the distribution of inter-sphere contacts is largely preserved at the RCP state, with frictionless particles possessing a mean number of 6 contacts~\cite{Bernal1960a, Donev2005, Alexander1998, Song2008, Briscoe2008, Torquato2010a}, well below the maximal 12 contacting spheres allowed in three dimensions \cite{Hales2005, Aste2008, Conway1997}. Despite these apparently ``universal" properties, to date the precise nature, or even existence, of a well-defined \textit{random-close packing} state is still subject of debate~\cite{Torquato2000, Kamien2007a, Parisi2010, Chaudhuri2010, Song2008}.

In this paper we consider the role of size asymmetry between particles in the structure of the RCP state.  Clearly, some measure of size distribution is unavoidable for any experimental system, and understanding the extent to which polydispersity modifies prediction of the monodisperse packing model remains an open question.  Beyond the inherent polydispersity of a single particle population, numerous material systems and processes, from ceramic powders to nanoparticle composites~\cite{German1989, McGeary1961, Furnas1931, Zheng1995, Newhall2012}, involve mixtures of intrinsically different particle populations of  distinct (mean) radii.  Unlike the RCP state of strictly monodisperse spheres, which is a fundamentally parameter-free problem in the infinite volume limit, the size and number ratios of constituent sphere populations introduce new parameters that have a critical influence on the structure of a randomly-packed mixture.  In this study, we consider the case of strictly {\it bidisperse} sphere packings to understand how the size ratio of spheres dictates the local and global structure of random-close packed mixtures, and more specifically, the {\it sensitivity} of that structure to variation in sphere composition.

\begin{figure}
%\centering  %\setlength\fboxsep{0pt} %\setlength\fboxrule{0pt}
\includegraphics[width=3.4in]{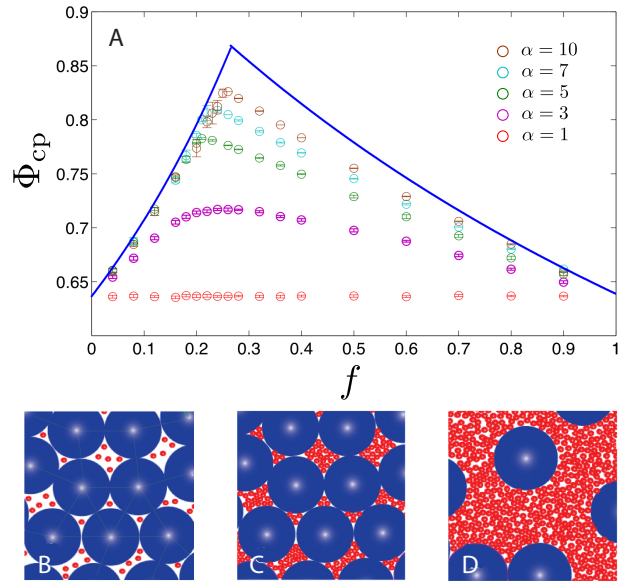}
\caption{\label{fig:1} Volume fraction, $\Phi_{\rm cp}$, of RCP binary sphere mixtures as a function of relative volume fraction, $f$, of small spheres for different radius (large/small) ratios, $\alpha$. The solid blue curve shows the prediction of IAL model for $\alpha\rightarrow\infty$ limit, eq. (\ref{eq: altoinfty}).  Circles show that maximum packing in simulated bi-dispersed mixtures for finite size asymmetry.  In (B-D), a schematic overview of the IAL model (small/large spheres are shown as red/blue):  (B) corresponds to a point on left solid curve where small spheres are loosely packed in the interstices of the jammed network of small spheres: (C) corresponds to the maximally-dense random binary jammed packing, with both small spheres closed-packed in the interstices of the close-packed large sphere network; and (D) corresponds to a point on right solid curve, where large spheres are ``dissolved" in the close-packed bulk of small spheres.}
\end{figure}

Bidisperse sphere mixtures are characterized by two dimensionless ratios:  the size ratio, $\alpha=\sigma_l/ \sigma_{s}\geq 1$ where $\sigma_l$, $\sigma_s$ are large ($l$) and small ($s$) sphere radii, respectively; and relative volume fraction of small spheres, $f= V_s/(V_l+V_s)$, where $V_i=(4 \pi/3) N_i \sigma_i^3$ is the volume occupied by $N_i$ spheres of type $i = s$ or $l$.  Experimental~\cite{Furnas1928,Yerazunis1965}  and simulation~\cite{Clarke1987,Kyrylyuk2010, Dodds1975, Richard1998, Hopkins2011} studies show that in random, bidisperse mixtures the volume fraction at close-packing $\Phi_{\rm cp}$ varies with $\alpha$ and $f$.  Namely, at fixed $f$, $\Phi_{\rm cp}$ generically increases with size asymmetry (as $\alpha$ increases or decreases from 1).  Moreover, at fixed $\alpha \neq 1$, $\Phi_{\rm cp}(f)$ is a non-monotonic function of $f$ with has a single maximum ($>\Phi^{mono}_{\rm cp} $) at compositions intermediate to the limiting monodisperse extremes at $f=0$ and $1$.  This generic behavior can be rationalized by a simple model, introduced by Furnas~\cite{Furnas1928, Westman1930, Zheng1995}, that considers random binary sphere packing in the extreme aspect ratio limit ($\alpha \to \infty$).  In this Infinite Asymmetry Limit (IAL) model, it is assumed that length scales of distinct components are so dissimilar that the random-close packing of each component decouples in the small- and large-$f$ limits, shown schematically in Fig.~\ref{fig:1}~B-D.  At low compositions of small spheres ($f < f_\infty$) the maximum density jammed structure is limited by the random-close packing of large spheres (occupying $\Phi^m_{\rm cp}\simeq 0.64$ of volume) while small spheres (unjammed) occupy the interstitial volume at a local volume fraction below the monodisperse RCP limit, $\Phi^m_{\rm cp}$ (see Fig.~\ref{fig:1}~B).  At larger small sphere compositions ($f \geq f_\infty$) it is not possible for small spheres to fill the interstitial volume without locally exceeding the RCP limit, hence, both small and large spheres are jammed but only small spheres form continuous, jammed networks with large spheres distributed, or ``dissolved," in the packing (see Fig.~\ref{fig:1}~D).  The total close-packing density in IAL has the following piece-wise continuous dependence on composition,
\begin{equation}
\label{eq: altoinfty}
\lim_{\alpha \to \infty} \Phi_{\rm cp} (f) = \left\{ \begin{array}{ll}  \Phi^{mono}_{\rm cp} /( 1- f) & f < f_{\infty}  \\ \Phi^{mono}_{\rm cp} /\big[f+   \Phi^{mono}_{\rm cp} (1- f) \big] & f \geq f_{\infty} \end{array} \right.
\end{equation}
The two branches of composition dependence meet at $f_\infty =(1-\Phi_{\rm cp}^m)/(2-\Phi_{\rm cp}^m) \simeq 0.26$ at which the small sphere population is locally close packed within interstitial volume of a close-packed network of randomly packed larger spheres.  The density reaches a maximum $\lim_{\alpha \to \infty} \Phi_{\rm cp} (f_\infty) =  \Phi_{\rm cp}^m + (1 - \Phi_{\rm cp}^m)\Phi_{\rm cp}^m \simeq 0.87$.   According to this IAL scenario, $f_\infty$ marks a singular cusp in $\Phi_{\rm cp} (f)$ in the limit of $\alpha \to \infty$ (see Fig.~\ref{fig:1}~A), indicating an abrupt transition in the jammed particle connectivity which we refer to as a {\it subjamming} transition:  for $f<f_\infty$, the subpopulation of small spheres are unjammed,  while for $f>f_\infty$, both sphere populations are jammed by neighbor contacts.

Fig.~\ref{fig:1}~shows the IAL prediction for $\alpha \to \infty$ behavior in comparison to simulations for finite $\alpha$ (method described below).   For finite $\alpha$, the evolution of $\Phi_{\rm cp}$ with small sphere composition is apparently smooth for nearly monodispere mixtures ($\alpha \gtrsim 1$), with the peak in $\Phi_{\rm cp} (f)$ becoming apparently sharper with increasing size asymmetry, suggesting an approach to the cusp at $f_\infty$ for $\alpha \to \infty$ predicted by the IAL model.  To date, previous studies of  random binary sphere packings have explored only modest degrees of asymmetry (e.g. $\alpha \leq 4$)~\cite{Richard2001, Kansal2002, Clusel2009, OHern2003}~or have not carefully sampled the composition axis at high size asymmetry~\cite{Farr2009, Kyrylyuk2010, Dijkstra1999, Hopkins2013},
%{\bf I don't know what this means}
 so the implications of a possible sub-jamming transition on the local and global structure of random bidispersed sphere packings at large $\alpha \gg 1$ remain largely unexplored.

In this study, we investigate numerical simulations of bidisperese sphere packings to explore how composition-dependence of RCP mixtures evolves from smooth, nearly monodisperse limit ($\alpha \gtrsim 1$) to the putative non-analytic behavior (marked by a discontinuous first derivative of $\Phi_{\rm cp}$) at $\alpha \to \infty$.  We envision two hypothetical scenarios that connect these limits:  (i) The structure of binary mixtures evolves smoothly and continuously with composition for all finite $\alpha$,  becoming strictly non-analytic in a singular $\alpha \to \infty$ limit; or alternatively, (ii) a subjamming transition, as a non-analytic dependence of packing on $f$, extends from $\alpha \to \infty$ down to a critical, but finite, value of asymmetry, $\alpha_c$.  Borrowing the language of thermodynamic phase transitions, in (i) a true ``transition" between small-sphere dilute and rich phases exists only at $\alpha \to \infty$, while for (ii) a line of first-order like transitions between two distinct ``phases" of packing extends from $\alpha \to \infty$ to a ``critical end point" at $\alpha_c$, below which structure evolves continuously from $f=0$ to $1$.  

\begin{figure*}
\includegraphics[width=6.6in]{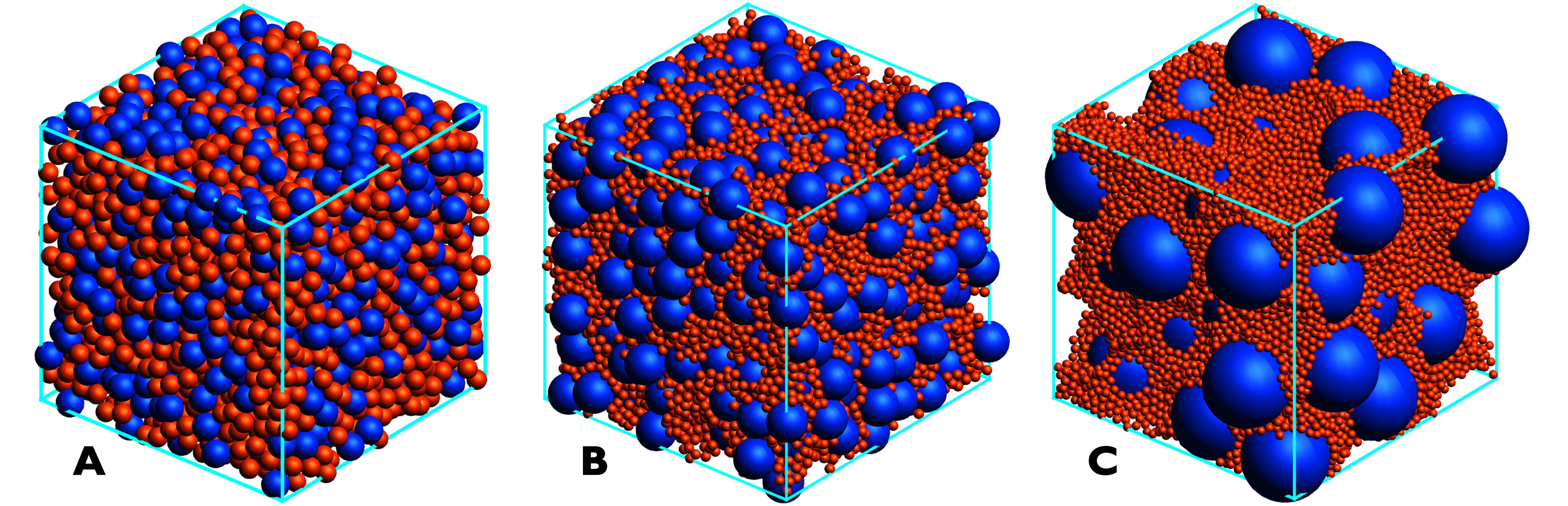}
\caption{\label{fig:2} Snapshots of three jammed binary sphere assemblies: nearly monodisperse (A) $\alpha$=1.4, $f$=0.5, N=5000, $\phi=0.6458$; moderately bi-disperse (B) $\alpha=4.0$, $f$=0.25, N=10000, $\phi=0.7532$; extremely bi-disperse (C) $\alpha=10.0$, $f=0.27$, N=18000, $\phi=0.8225$}
\end{figure*}

To distinguish between these two scenarios, we perform numerical simulations of athermal, jammed binary sphere packings over the full composition range and in the range of asymmetries $1\leq\alpha\leq10$, and compare several measures of local and global structure of random-close packed binary mixtures.   We report evidence for scenario (ii), with a critical value of asymmetry $\alpha_c \simeq 5.75$ above which structural measures of packing become sharp (i.e. apparently non-analytic) functions of sphere composition along a line of sub-jamming transitions extending to $\alpha \to \infty$ and $f \to f_\infty$.   Motivated to understand the appearance of abrupt transitions at finite $\alpha$, we analyze the ``cooperativity" of small sphere jamming for small $f$ within the disjoint interstitial volumes between jammed (or nearly jammed) large spheres, and based on this, construct a heuristic argument to estimate the value of $\alpha_c$.  Overall, these results illustrate a highly, non-trivial dependence of RCP structure on size asymmetry of constituent elements that may have important practical implications for engineering mesoscale structure in random particle mixtures.

This article is organized as follows.  In section II, we briefly summarize the numerical simulation protocol for binary sphere mixtures and define three measures of packing structure:  closed packing density, $\Phi_{\rm cp}$; rattler fraction, $R_s$; and ``contact inhomogeneity", $Q$, introduced in \cite{Richard1998} as a statistical measure of local contacts between like and unlike spheres.  In section III, we present numerical maps of $\Phi_{\rm cp}$, $F_s$ and $Q$ over the full composition range and for $\alpha \leq 10$, and we further analyze the emergent discontinuity of rattler fraction via finite size scaling to characterize the value of and power-law approach to the critical end point, $\alpha_c$.  In section IV, we present a heuristic argument to understand emergence of a sub-jamming transition at a finite value of $\alpha$, and further substantiate this by statistical analyses of spatial correlations among distinct spatial regions of small-sphere jamming.  We conclude with a brief discussion of transitions in long-range connectivity within the small and large-particle contact networks, and potential implications of the ``sub-jamming" transition for random mixtures of particles of distinct shape.

\section{\label{sec:nummeth}Simulation and Analysis of Random-Close Packed, Binary Spheres}
\subsection{Binary Packing Protocol }
\label{subsec:proto}

We simulate random, close-packed mixtures of athermal, frictionless spheres based on the protocol for monodisperse, jammed spheres of O'Hern {\it et al} \cite{OHern2003}.  For a given composition $f$, size ratio $\alpha$, and total number of particles, $N$, we specify the numbers, $N_s$ and $N_l$, of small and large spheres, respectively, by truncating $N_l$ to the nearest integer value and choosing $N_s=N-N_l$. For $\alpha=1$, this truncation alters the final composition by $\delta f=1/N \approx 10^{-4}$. The centers of all spheres are randomly distributed within a three-dimensional periodic simulation box.  Initially, the radii of these particles are randomly assigned values of $\sigma_s$ and $\sigma_l$, according to the fixed values of size asymmetry, $\alpha$ and small-sphere composition $f$,
\begin{equation}
\label{eq: alf}
\alpha =\sigma_l/\sigma_s; \ f = \frac{N_s \sigma_s^3}{ N_s \sigma_s^3 + N_l \sigma_l^3} ,
\end{equation}
at an initial volume fraction far below the onset of inter particle overlap ($\Phi_{\rm init} = 10^{-2}$). We choose an initial value of composition, $f$ then for a fixed total number of spheres, $N$, calculate nearest integer values of Ns and   From this point, the density of the distribution is increased in stepwise intervals of $\delta \Phi_0 = 10^{-3}$ by increasing all particle radii by $(\delta \Phi)^{1/3}$ with particle centers fixed.  After each density increase, the positions of the particles are relaxed until no particle pairs overlap.  The relaxation is achieved via forces generated by soft-repulsive potentials between particles of type $i$ and $j$ at a center-to-center distance $r$
\begin{equation}
V_{ij}(r) = \left\{ \begin{array}{ll} \frac{\epsilon}{2} (1 -  r/\sigma_{ij} )^2 ,& r \leq \sigma_{ij}  \\
0, & r > \sigma_{ij}  \end{array} \right.
\end{equation}
where $\sigma_{ij} = \sigma_i + \sigma_j$ and $\epsilon$ is an energy scale.  Overlapping particle positions are relaxed via a quasi-Newton (L-BFGS) method, until the total energy falls below tolerance $10^{-10}\epsilon$ at each density iteration. This process continues until a local minimum with a non-zero energy (above this tolerance) is found, indicating non-zero overlap in the configuration and that the packing is beyond the jamming point. From this point, the algorithm reverts to the nearest pre-jammed configuration, and proceeds again with a reduced density increment $\delta\Phi_n=\delta\Phi_{n-1}/2$, where $n$ is the number of times the jamming threshold is exceeded before terminating. This continues iteratively until a state within $\delta\Phi_n \leq 10^{-5}$ of an unjammed configuration is reached, a point we take to be the onset of jamming.  

Below we describe results of this algorithm for binary sphere mixtures in the range of asymmetry $1 \leq \alpha \leq 10$ and for total particle numbers $N$ ranging from $5000$ for nearly monodisperse case to $18000$ for high size asymmetry ($\alpha \geq 8$). Simulations with size ratio $\alpha \approx 6$ were carried out with $N=14000$.  Simulations are carried out over the full range of $0\leq f \leq 1$ in coarse intervals of  $\delta f = 0.05$ at low- and high-$f$ and at finer intervals $\delta f = 10^{-3}$ in the range of $f$ going from $0.2 - 0.25$ to capture the sharp evolution of jammed structure in very asymmetric mixtures and $\delta f = 0.1$ for $f\geq 0.3$ where properties of the exhibit a more gradual composition dependence.  For most points in $\alpha-f$ phase space, 20 independent simulations are performed to generate statistics. Fig \ref{fig:2}A-C show snapshots of jammed binary sphere configurations for nearly monodisperse ($\alpha=1.4$), moderately bidisperse ($\alpha=4.0$), and extremely bidisperse ($\alpha=10.0$) cases

Once the jammed configuration is obtained, we analyze the distribution of contacts in the configuration, and decompose the packing into jammed and unjammed spheres, or ``rattlers".  Contacts are assigned with some tolerance $\sigma_{tol}=10^{-7}$ for sphere pairs of separation within $\sigma_i + \sigma_j + \sigma_{tol}$.  In order for a sphere to be jammed, it must have no translational degrees of freedom, that is, the sphere center lies inside a polyhedron formed by joining centers of its contacts and the jammed sphere must have at least four contacts not all in the same hemisphere. Our configurations are {\it locally jammed} under the classification scheme described in Ref.~\cite{Donev2004}. If this criterion is not met, the sphere is identified as a rattler and its contacts are removed from the network of force supporting contacts in the packing. Each contacting sphere is checked for rattlers and the procedure is repeated until no further rattlers can be removed from the contacting structure.

\begin{figure*}
\centering  %\setlength\fboxsep{0pt} %\setlength\fboxrule{0pt}
\includegraphics[width=7.0in]{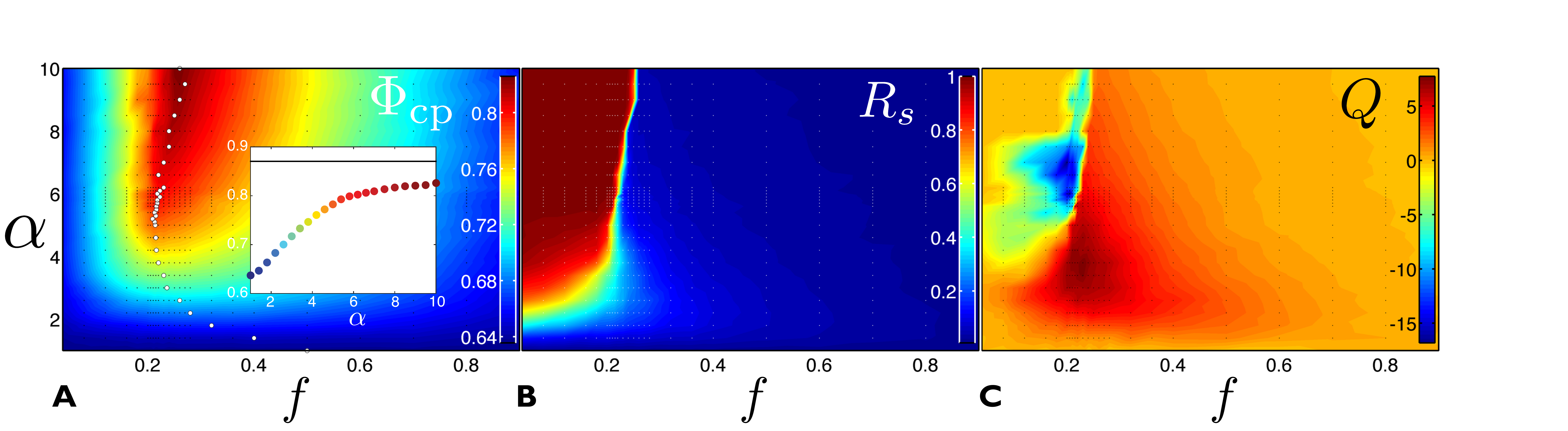}
\caption{\label{fig:3} Plots showing three structural properties of binary jammed sphere packings as a function of radius ratio, $\alpha$ and relative small sphere volume composition, $ f$. In (A) total occupied volume fraction, $\Phi_{{\rm cp}}$, of simulated packings increases with size ratio, with filled white circles corresponding to composition yielding maximum packing fraction at each $\alpha$, and the inset showing the asymptotic approach of maximum close packing density to IAL (solid line) with increasing size ratio.  In (B), the rattler fraction of small spheres, $R_s$.  In (C),  the contact inhomogeneity $Q$ as defined in eq. (\ref{eq: Q}). Simulated points are shown as white points in the $\alpha-f$ plane, and values of measured parameters are interpolated according the color scales shown.  }

\end{figure*}

\subsection{Structural Analysis of Packings}
\label{subsec:obs}

We analyze each simulated packing in terms of three distinct structural properties:  global volume occupancy; distribution of load-bearing (or jammed) spheres; and the statistics of contact between like and unlike spheres.  Clearly, the ratio of occupied to total volume, or the close-packing density, $\Phi_{\rm cp}$, serves as natural metric of the first property.  For the second property, we analyze the rattler fraction, 
\begin{equation}
R_s \equiv N_s({\rm rat})/N_s ({\rm tot})
\end{equation}
where $N_i({\rm rat})$ is the total number of ``rattlers" (type $i$) removed from the contact algorithm described above from a structure including $N_i ({\rm tot})$ spheres.  For the third property, we adopt the analysis of Richard and coworkers \cite{Richard1998,Richard2001} to determine the degree of local segregation of contacts between large and small spheres in the jammed network.  We call this quantity ``contact inhomogeneity," $Q$, and define it to be the excess fraction of {\it unlike} (large-small) contacts relative to a purely random and spatially uncorrelated network of same total number of contacts with the same numbers of contacting large and small spheres (e.g. nodes in the contact network).  All contact statistics are calculated after removal of ``rattlers" from the jammed configurations.  Consider a jammed network composed of  $n_{l}$ total large sphere contacts and $n_{s}$ total small sphere contacts (e.g. the number of large spheres in the contact network is $n_{l}/z_l$ where $z_l$ is the mean number of contacts per large sphere). Randomly distributing $N_c$ total inter-particle contacts among large- and small-sphere contacts, we arrive at a probability (per contact) of unlike sphere contact of $2 n_s n_l/(n_l +n_s)^2$.  The actual fraction of unlike contacts in a contact structure is determined from the numbers of large-large ($\eta_{ll}$),  small-small ($\eta_{ss}$) and small-large ($\eta_{sl}$) contacts.  The contact inhomogeneity, $Q$, is defined as the difference between the actual unlike contact fraction, and the maximally-random fraction of unlike contacts that would occur in perfectly uncorrelated like/unlike contact network,
\begin{equation} 
\label{eq: Q}
Q \equiv \frac{\eta_{sl}} {\eta_{ll}+\eta_{sl}+\eta_{ss}} - \frac{2 n_s n_l}{(n_l +n_s)^2} .
\end{equation}
From this definition, a positive (negative) value of $Q$ corresponds to the relative excess (deficit) of contacts between large and small spheres, indicating some measure of enhanced local mixing (segregation) between unlike sphere populations as compared to a purely random distribution of inter-particle connections of the same number.

Fig. \ref{fig:3}A shows a map of the mean {\it close packing fraction} $\Phi_{cp}$ over the range of simulated asymmetric packings, $1 \leq \alpha \leq 10$.  As is summarized in the introduction and Fig.  \ref{fig:1}, this map shows that 1) for fixed composition $f$,  $\Phi_{cp}$ increases with radius asymmetry, and 2) for fixed asymmetry $\alpha$, close packing fraction increases from the monodispere limits at $f=0$ and $1$ to a single maximum at an intermediate composition value.   The inset of Fig.~\ref{fig:3}A shows the increase of the maximum-composition value of $\Phi_{cp}$ for increasing $\alpha$, consistent with an asymptotic approach to maximum value of 0.87 predicted by the IAL model in the $\alpha \to \infty$ limit.   For small to moderate $\alpha$, when two species are comparable in size, density varies gradually with increasing $f$. At large $\alpha$, this dependence of $\Phi_{cp}$ on $f$ becomes sharper near its maximum, although it is not possible to resolve any evidence of discontinuity in the slope of $\Phi_{cp}(f)$ due to the finite resolution of composition increments.  Fig. \ref{fig:3}(A) also shows the dependence of locus of maximal-$\Phi_{cp}$ compositions on different $\alpha$.  For relatively symmetric mixtures ($1 \leq \alpha \lesssim 5-6$), maximal-$\Phi_{cp}$ compositions shift with increasing $\alpha$ to smaller compositions from the limiting $f=0.5$ case of $\alpha =1$.  This is followed by a shift back towards higher $f$ for $\alpha \gtrsim 5-6$. 

Fig. \ref{fig:3}B shows a map of the {\it small-sphere rattler fraction}, $R_s$ of simulated packings, showing generically that the fraction of unjammed small spheres {\it increases} with size asymmetry for a given $f$, and for a fixed asymmetry, decreases with small sphere composition.  This behavior is consistent with the intuitive notion that with decreasing radii, small spheres can better avoid contact in a globally jammed packing.  Not unlike the behavior of $\Phi_{cp}$, the evolution of $R_s$ in small-$f$ to large-$f$ mixtures is gradual for modest size asymmetry, $1\leq \alpha \lesssim 5-6$.  But for highly asymmetric $ \alpha \gtrsim 5-6$ the dependence of $R_s$ on $f$ becomes sharp and first-order like, dropping rapidly from 1 to 0 over a vanishing composition range, coincident (or nearly so) with the fixed-$\alpha$ maxima in $\Phi_{\rm cp}(f)$ at highly asymmetric $\alpha$.

Fig. \ref{fig:3}C shows the map of the {\it contact inhomogeneity}, $Q$, or the statistical excess of unlike contacts between jammed particles, showing a complex behavior in the $f-\alpha$ plane.  At low $\alpha$ as well as intermediate to large-$f$, $Q$ is positive, indicating a propensity for mixing of contacts unlike size particles.  At high $\alpha$ and low $f$, $Q$ is negative indicating a tendency for demixing between large and small contacts.  As sketched in the introduction for highly asymmetric mixtures, $Q >0$  is consistent with the ``dissolution" of large sphere contacts in the predominantly small sphere jammed network at large $f$ while $Q <0$ is consistent with the concentration of jammed small spheres in the interstices between large spheres for sufficiently low $f$.  Notably, we observe an abrupt transition between negative and positive values of $Q$ with increasing $f$ for high asymmetry mixtures, again appearing for $\alpha\gtrsim 5-6$, while $Q$ evolves smoothly with $f$ below this value.

While the packing fraction, (small sphere) rattler fraction and contact inhomogeneity each exhibit qualitatively different dependencies on sphere composition $f$, taken together the behaviors mapped in Fig. \ref{fig:3} show that the evolution from small-sphere poor to small sphere rich becomes increasingly sharp with increasing $\alpha$, tending towards a qualitatively different behavior for $\alpha \gtrsim 5-6$.  The coincident rapid jumps in $R_s(f)$ and $Q(f)$ along the locus of maximal $\Phi_{\rm cp} (f)$ for large $\alpha$ (in contrast to the smooth evolution with $f$ for small $\alpha$) suggest an abrupt transition in the jammed network structure for large, but finite asymmetries.  This scenario is consistent with non-analytic dependencies of structural quantities on $f$ (i.e., discontinuities in order parameters or their derivatives).  To analyze the distinction between small- and large-$\alpha$ behavior we consider the small rattler susceptibility 
\begin{equation}
\chi_{R_s} \equiv -\bigg(\frac{\partial R_s}{\partial f}\bigg)_{\alpha} ,
\end{equation}
which characterizes rate at which small spheres are incorporated in the jammed particle network when the small sphere composition is increased.  For simulations carried out on a finite set of compositions $f$, we use the finite difference approximation $\chi_{R_s}\simeq [R_s(f+\delta f)-R_s(f)]/ (\delta f)$.  For given number of spheres $N=N_s+N_l$ in the simulation box, there is, from eq. (\ref{eq: alf}), a lower limit to the composition increment $(\delta f)_{\rm min} = f N_s^{-1}(1-f +f \alpha^3) \sim 1/N$ where simulation increments differ only by the inclusion of $\pm$ a single small and large sphere.  In practice, for the total $N$ values used, we carry out, across the locus of maximal $\Phi_{\rm cp}$ in composition, increments of $ (\delta f) =0.05$ (coarse resolution) or $0.002$ (finer resolution) 
, which places a resolution limit on the magnitude of discrete approximation of rattler susceptibility when it approaches the limiting upper value $(\delta f)^{-1}$.

\begin{figure}
\includegraphics[width=3.3in]{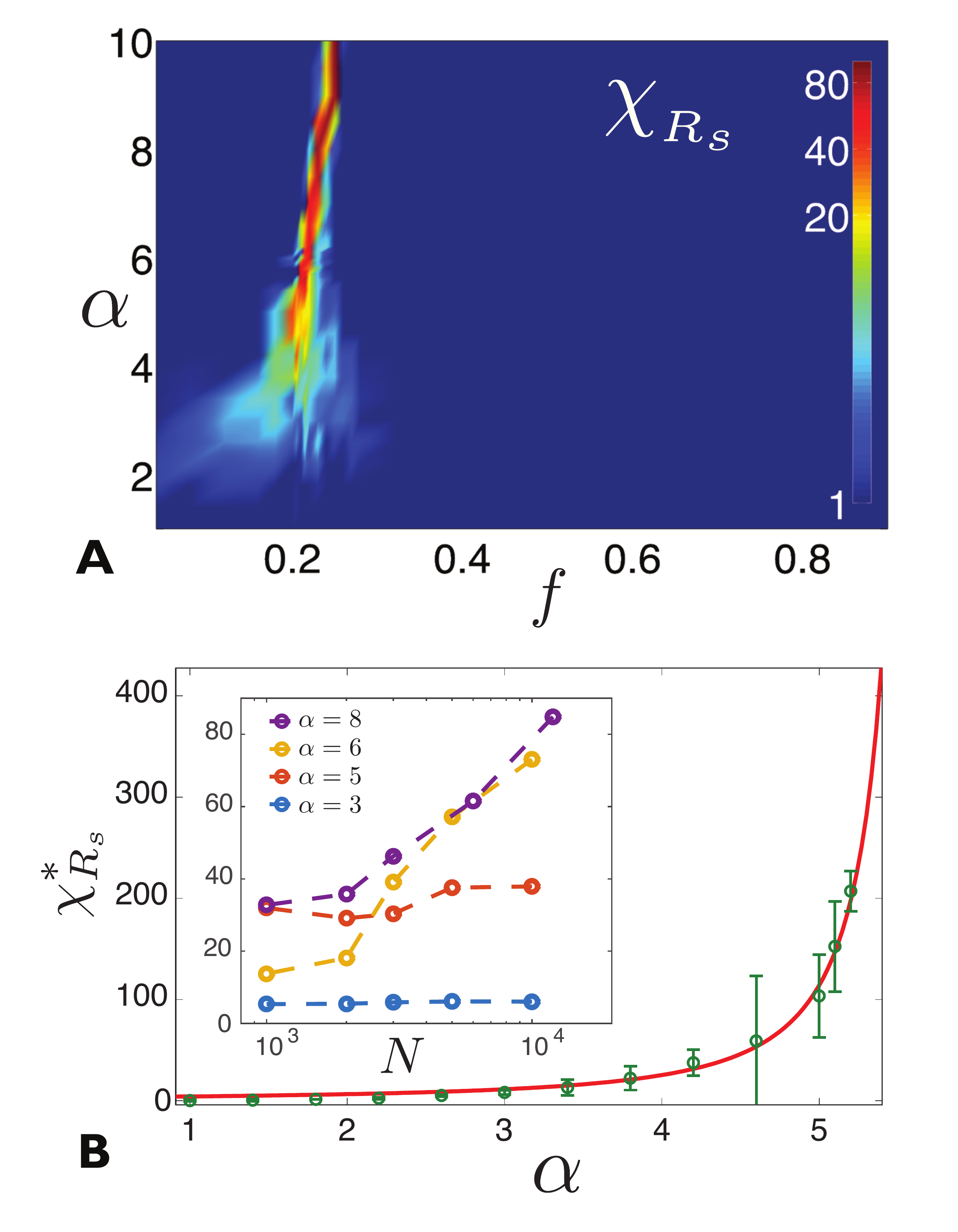}
\caption{\label{fig:4} 
(A) Density plot of peak rattler susceptibility, $\chi^*_{R_s}(\alpha)={\rm max}_f\big[\chi_{R_s}  (f,\alpha) \big]$ as a function of size ratio, $\alpha$ (note logarithmic color scale). (B) Plot showing divergence of rattler fraction susceptibility with increasing size ratio $\alpha$. Red curve fits numerical data to a power law, $\chi_{\rm F_r max}=A/(\alpha_c - \alpha)^\gamma$ yielding $\alpha_c\simeq5.75$ and $\gamma\simeq1.78$. (inset) shows dependence of maximum susceptibility on total number of spheres in the system.  }
\end{figure}

In Fig \ref{fig:4}A, we plot a map of the small rattler susceptibility $\chi_{R_s}$ in the full $f-\alpha$ plane, constructed via coarse composition increments, $\delta f = 0.05$ and for total sphere numbers $N =5000-18000$.  For small size asymmetry, the susceptibility remains of order unity, consistent with the continuous and smooth evolution of $R_s(f)$ with small sphere composition.  In the high asymmetry regime, $\alpha \gtrsim 6$,  the susceptibility becomes peaked along the locus of maximal close-packing compositions, yielding a value of order of the coarse-resolution limit (i.e., $\chi_{R_s} \approx (\delta f)^{-1}$), consistent with a discontinuous decrease in the number of small sphere rattlers across this line.  In the inset of Fig \ref{fig:4}B, we plot the peak rattler susceptibilities, $\chi^*_{R_s}(\alpha)={\rm max}_f\big[\chi_{R_s}  (f,\alpha) \big]$, as a function of increasing total sphere number ($N \simeq 10^3-10^4$) for size asymmetries $\alpha=3,5,6$ and $8$.  For the two lower asymmetries ($\alpha = 3$ and $5$) we observe that peak susceptibilities are independent of total system size, suggestive that gradients of $R_s(f)$ remain finite in the $N \to \infty$ limit.   In contrast, for the more asymmetric values  ($\alpha = 6$ and $8$), $\chi^*_{R_s}(\alpha)$ continues to increase with $N$, suggestive of a first-order like discontinuity in $R_s(f)$ that is smeared-out by finite-$N$ and $\lim_{N \to \infty} \chi^*_{R_s}(\alpha) \to \infty$ in high-asymmetry mixtures.

We probe the nature of transition between this continuous and apparently discontinuous evolution of $R_s$, by analyzing the increase in $\chi^*_{R_s}(\alpha)$ as $\alpha$ increases from the monodisperse case towards the high-asymmetry regime.  Here, to extend resolution to higher range of $\chi_{R_s}$ (and larger $\alpha$) we carry out simulations over a finer composition grid, with $ (\delta f) =0.002$.  Fig \ref{fig:4}B, plots the peak resolution $\chi^*_{R_s}(\alpha)$ for $\alpha=1-5.2$ (at higher $\alpha$ the discrete approximation exceeds $50\%$ of its resolution limited value, $(\delta f)^{-1}$, and/or exhibits the finite-$N$ dependence shown in Fig.  \ref{fig:4}B inset).  Over the range $\alpha \simeq 3.4 - 5$, $\chi^*_{R_s}(\alpha)$ grows very rapidly from order unity to an apparent divergence at a finite value of $\alpha$.  To estimate a critical value for onset of truly discontinous behavior of $R_s(f)$, we fit the $\chi^*_{R_s}(\alpha)$ data to a power-law divergence,
\begin{equation}
\chi^*_{R_s}(\alpha) \simeq \frac{\chi_0}{|\alpha_c-\alpha|^\gamma},
\end{equation}
whose best fits yield an exponent $\gamma=1.78\pm 1.4$ and critical size asymmetry,
\begin{equation}
 \alpha_c=5.75\pm 0.8
 \end{equation}
Given the limit range of $\chi^*_{R_s}(\alpha)$ and $|\alpha_c-\alpha|$ achievable with finite resolution limits imposed by $\delta f$ and $N$, we are not in a position to strictly verify the form of the divergence (i.e., as power law vs. other functional forms).  We nevertheless take the apparent divergence in $\chi^*_{R_s}(\alpha)$  as $\alpha \to \alpha_c$, as evidence of a finite and critical value of size asymmetry beyond which the transformation from small-sphere poor to small-sphere rich jammed packings will become truly singular (accompanied by discontinuities in structural measures or their derivatives) in the $N \to \infty$ limit.  That is, $\alpha= \alpha_c \simeq 5.75$ and $f\simeq0.214$ has some hallmarks of a {\it critical end point} of a line of first-order transitions that extends from this finite value of size asymmetry to $\alpha \to \infty$.  We denote the location of the discontinuity in packing structure as $f_{sub}(\alpha)$, and associate the discontinuity with a {\it sub-jamming transition} of the small spheres within the globally jammed binary packing that becomes truly sharp (i.e., singular in the sense of non-analytic dependence of measures of stucture on $f$ at $f_{sub}(\alpha)$) only above a critical asymmetry, $\alpha \geq \alpha_c$.

 \subsection{ Cooperativity in Small Sphere Jamming}
 
 \label{sec: cooperativity}

Motivated by the apparently discontinuous structural transition of jammed packings and its connection to the abrupt change in the population of small-jammed spheres along the line $f=f_{sub}(\alpha)$ for $\alpha \geq \alpha_c$, we analyze the statistics of small sphere jamming in more detail.  In particular, we focus on the spatial correlations of small sphere jamming from the point of view that, in general, large spheres are jammed (i.e., not rattlers) in all globally jammed packings, but small spheres may or may not be ``sufficiently confined" in the binary packing to jam with a high probability.  For example, in the limit of very large $\alpha$, the IAL scenario suggests that, at low $f$, small spheres rattle around in the interstices of jammed large spheres until they reach a sufficient total number to fill those regions at close packing, at which point they must also be jammed.  

Here, we explore this scenario by decomposing space in the jammed packings into disjoint volumes corresponding to ``voids" between large spheres in the packing.  Specifically, large-sphere voids are Delaunay simplices (i.e. tetrahedra) defined by the dual partition of the Voronoi tessellation generated by the large-sphere centers.  Disordered monodispere packings are often characterized in terms of the structure and correlations of the polytetrahedral coordination defined by this graph \cite{Anikeenko2007, Anikeenko2008, Charbonneau2013a}.  Our analysis uses large sphere centers as sets of points to uniquely define partitioning of space; small sphere centers fall within one of these tetrahedral domains whose vertices are defined by large-sphere centers, hence only a fraction of the tetrahedral volume will be accessible to small spheres due overlap constraints with the vertex spheres. 

\begin{figure}
\includegraphics[width=3.5in]{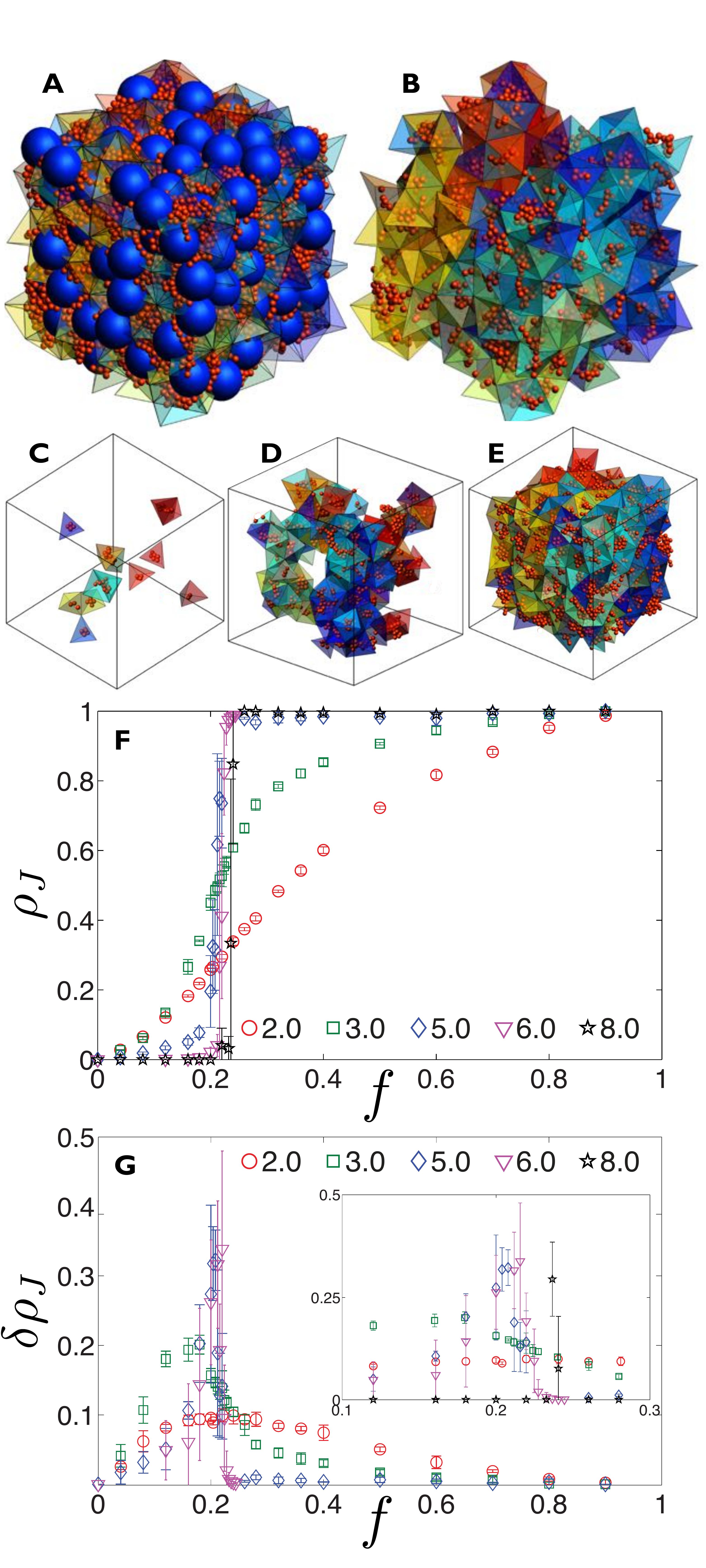}
\caption{\label{fig:5} (A) Snapshot showing all the spheres of a jammed configuration with $\alpha=6$, $f=0.24$. Positions of small spheres assigned to enclosing tetrahedral (simplical) domains whose vertices are large sphere centers.  In (B), the same configuration showing only jammed small spheres and their corresponding simplifies (face colors are randomly assigned for visibility). (C-E) Shows the rapid evolution of small sphere jamming at $\alpha=6$ with increasing $f$ from (C) $f=0.21$ to (D) $f=0.23$ to (E) $f=0.25$, with spatially correlated simplifies visible at intermediate fractions of jammed small spheres.  In (F) fraction of jammed interstitial volumes $\rho_J$ and, in (G), excess jamming probability for simplices with jammed neighbors, $\delta \rho_J$, both plotted as functions of small sphere composition, and for increasing values of size asymmetry.}
\end{figure}

Based on this simplicial decomposition of the binary jammed packings, we assign an Ising-like variable $S_{I}$ to each simplex $I$, such that $S_{I}=0$ encloses no jammed small spheres, otherwise $S_{I}=1$ if it encloses one or more jammed small spheres.  From this the total fraction of jammed interstitial volumes is
\begin{equation}
\rho_J=n_{\rm sim}^{-1} \sum_{I=1}^{n_{\rm sim}} S_{I},
\end{equation}
where $n_{sim}$ is the total number large-sphere simplices.  A plot of $\rho_J$ as function of increasing small-sphere composition is shown in Fig.  \ref{fig:5}F for range of size asymmetries, varying from $\alpha =2$ (nearly monodisperse) to $8$ (high asymmetry). For low asymmetry $\alpha < \alpha_c \simeq 5.75$, $\rho_J$ increases continuously from 0 to 1 as $f$ increases, with a maximum slope that grows with $\alpha$, while above the critical size asymmetry, we find an apparent discontinuous jump in $\rho_J$ at $f_{sub}(\alpha)$, consistent with the sub-jamming scenario analyzed above.  

To investigate the distinction between smooth and sharp increases of $\rho_J$ with $f$ we analyze local correlations of small sphere sub-jamming in neighboring simplicial domains.  Specifically, we explore the question, are small spheres within interstitial volume of large-spheres more or less likely to be jammed, if the small spheres in the neighboring interstitial region are jammed? and if so, under which conditions?  We define $\delta \rho_J$ to be the excess likelihood of an interstitial region to be jammed if at least one its neighboring region is jammed.  This is determined from operator $N_I$ which is 0 if $S_J=0$ for all four neighbors $J$ of $I$, or 1 if at $S_J=1$ for at least one neighbor (i.e. $N_I=1$ if any neighbor regions are jammed).  From these we find that 
\begin{equation}
\delta \rho_J =  \frac{ \sum_{I=1}^{n_{sim}} S_I N_I}{\sum_{I=1}^{n_{sim}} N_I}- \rho_J ,
\end{equation}
where the first term is the fraction of interstices with jammed neighbors that are themselves jammed.  We plot $\delta \rho_J$ as function of small-sphere composition, for the same range of $\alpha$ in Fig  \ref{fig:5}G.  Here we see $\delta \rho_J$ is generically maximal at an intermediate value of $f$, with a maximal peak that increases in amplitude and sharpness with increasing $\alpha$.  For example, for the nearly monodisperse case of $\alpha=2$, $\delta \rho_J$  increases to a broad, maximal excess probability of jamming of about 0.1, while for the most asymmetric case of $\alpha = 8$, we find that $\delta \rho_J$ only increases substantially in a narrow range near to $f_{sub}(\alpha=8)$, but it rises to a much larger enhancement of interstitial jamming approaching 0.5 at the maximum.   

This trend indicates that small-sphere jamming in the disjoint domains defined by large-sphere interstices becomes more {\it cooperative} as the size asymmetry, $\alpha$, increases.  This observation is consistent with the fact that for small $\alpha$ cooperativity remains low, and thus, large sphere voids jam statistically independently of one another as small-sphere number grows, leading to the gradual rise of $\rho_J$ with $f$.  Conversely, for $\alpha \geq \alpha_c$ cooperativity between neighboring interstices is sufficiently strongly enhanced near the sub-jamming transition such that sub-jamming occurs in a ``all or nothing", discontinuous fashion as $f$ reaches $f_{sub}(\alpha)$.   In short, the increase of local correlations of small sphere jamming in neighbor voids of large spheres with $\alpha$ is consistent with continuous ($\alpha  < \alpha_c$) vs. discontinuous/first-order ($\alpha > \alpha_c$) suggested by the analysis of the previous sections.  In the following section, we discuss the possible geometric origin of cooperative sub-jamming behavior.

\section{\label{sec:Disc} Discussion}

The results described above suggest that randomly jammed packings of binary spheres can be divided into two distinct classes.  For weakly asymmetric sphere sizes $\alpha < \alpha_c \simeq 5.75$, structural properties (e.g., packing fraction and contact structure) of the mixture evolve continuously with small sphere composition $f$.  Above this critical asymmetry $\alpha \geq \alpha_c$, packings can be divided into two ``phases" separated by a sharp, sub-jamming transition at $f=f_{sub}(\alpha)$:  a state of incomplete small-sphere jamming and a state where both populations of spheres are nearly completely jammed.  Motivated by these observations, here we discuss both the possible geometric mechanisms underlying this behavior, as well as the implications for the composition dependence of global connectivity properties of small-sphere networks in binary mixtures.

\subsection{\label{sec:crit} Geometric Origins of Sub-Jamming Transition}

The simulations in Sec. \ref{subsec:obs} argue that the scenario suggested by the IAL model predicted on the $\alpha \to \infty$ limit holds for range of finite $\alpha>\alpha_c$.  In particular, this scenario is predicated on the {\it decoupling} of large-large and small-small jamming in the small-$f$ regime, where large-spheres form a RCP network among themselves and the loose population of small spheres are confined (largely as rattlers) to the large sphere interstices.  While this structural picture is somewhat intuitive, identifying the specific mechanisms underlying this decoupling behavior remains a challenge, given the non-equilibrium (and potentially protocol dependent) pathway to the randomly-jammed state of binary spheres.  That is, why, for sufficiently small $f$ and large $\alpha$, small spheres are simply ``swept aside" into the interstitial regions between large spheres upon compression, to the extent that large spheres jam only when they are jammed by other large spheres?  What prevents small spheres from establishing enough large-sphere contacts to perturb large sphere jamming in this regime?  

\begin{figure}
\includegraphics[width=3.4in]{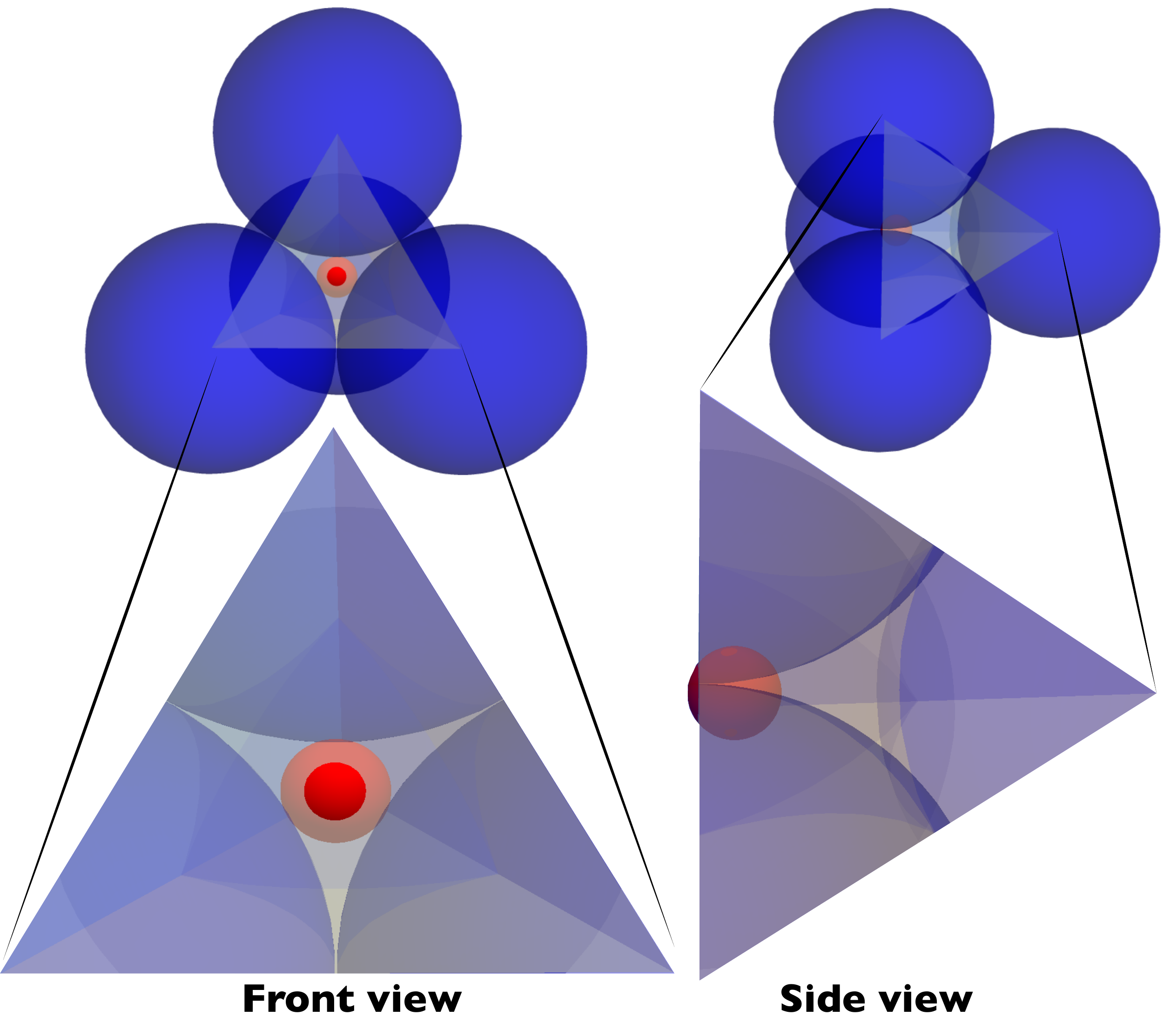}
\caption{\label{fig:6} Schematic of the geometry underlying the ``throttling criteria", or the critical size ratio that allows transmission of small-sphere contacts through interstitial voids of close packed large spheres. A small (red) sphere is considered within the intersitices of the close-packed large (blue) spheres, tetrahedral packing.  Here, the small sphere is shown slightly protruding through the volume defined by the simplex of large sphere centers (i.e. slightly beyond the ``throttling criteria"), with faces shown as transparent white.}
\end{figure}

One element of this behavior can be associated with the fact that while large spheres can in general have sufficient small-sphere contacts to be jammed by small spheres, the converse is not in general true, as number of large spheres that can contact a given small sphere is highly limited.  This latter property can be quantified by the generalized ``kissing number", $z_{s/l}(\alpha)$, the maximum number of non-overlapping large spheres that can contact a small sphere.    This number is known to be related to the Tammes problem~\cite{Conway2013}, or the maximum geodesic radius, $\rho_{max}=\sigma_s \sin^{-1}\big[1/(1+\alpha_{max}^{-1} )\big]$, of $z_{s/l}$ non-overlapping discs on the surface of a sphere.  Notably, the maximal $\alpha_{max}(z_{s/l})$ for which the small-sphere may possess either $z_{s/l}=4$ or $z_{s/l}=6$ contacts are known to be determined by, respectively, tetrahedral or octahedral arrangements of large spheres around a central sphere, from which it can be shown that $\alpha_{max}(z_{s/l}=4)=4.45$ and $\alpha_{max}(z_{s/l}=6)=2.41$.  Therefore, above these respective size ratios, large sphere contacts are not able to provide the minimal number of contacts for local jamming of small sphere in the first case, or the mean number of contacts required for isostaticity of jammed small spheres in the second.  Clearly, small spheres may also be jammed by other small spheres or combinations of large and small sphere contacts, but these basic geometric arguments suggest that, as $\alpha$ increases and the fraction of small spheres is low (small $f$), there is insufficient space around small spheres for the contacts required by small sphere jamming to be achieved large spheres alone, consistent with the observation that small-spheres ``decouple" from the jamming of large spheres in this limit.

While the forgoing arguments are consistent with emergence of interstitial ``loose-packing" of small-spheres in the large-$\alpha$, low-$f$ regime, there remains the question, what sets the scale of the critical asymmetry, $\alpha_c$?  The observation of critical-end point like behavior at $\alpha_c \simeq 5.75$ occurs at a somewhat higher value of asymmetry than, for example, the size ratio $\alpha_{max}(z_{s/l}=4)=4.45$ above which one or more small sphere can fit into a locally closed-packed tetrahedra of large spheres.  An alternative mechanism to set this threshold is associated with the cooperativity of small-sphere jamming in separate disjoint interstitial volumes between jammed large spheres, as analyzed in Sec. ~\ref{sec: cooperativity}.  While for $\alpha>\alpha_{max}(z_{s/l}=4)$ small spheres are in principle small enough to be unjammed between locally-dense large spheres, they are in general not necessarily small enough that contacts between jammed small spheres can pass between neighboring interstitial volumes.  Fig. \ref{fig:5}F-G suggests that for $\alpha \geq \alpha_c$, the sub-jamming transition occurs cooperatively, with nearly all interstitial volumes becoming jammed along a critical line $f_{sub}(\alpha)$.  Thus, a sufficient geometric condition for this cooperative interstitial jamming is that small sphere contacts pass from one interstitial volume to its neighbor.  We estimate this ``throttling criteria" by considering size ratio where a small-sphere can protrude through the planar face of a tetrahedrally-close packed volume defined by four contacting large spheres (see schematic in Fig.  \ref{fig:6}).  Based on this simplified geometry, for $\alpha > \alpha_{throt} = 6$ we estimate that small spheres are able to pass contacts between neighboring interstitial regions.  Of course, in a RCP network of large spheres the simplicial volumes are at least somewhere looser, and polydisperse, than the tetrahedral-close packed one, and therefore, it is expect that $\alpha_{throt} = 6$ should really be taken as an upper bound to the size ratio at which interstitial small-sphere jamming becomes highly cooperative with neighbor regions.  Notwithstanding the limitations of this crude geometric argument, this estimate is reasonably consistent with the value of $\alpha_c \simeq 5.75$~\footnote{Note that the size ratio at which small spheres are able pass freely between the triangular faces of tetrahedrally-closed packed large spheres, $\alpha_{pass} \simeq 6.46$ is somewhat larger than $\alpha_{throt}=6$.} estimated from the critical-end point like behavior observed in the small-sphere rattler population.

\subsection{\label{sec: connect} Sub-Jamming and Global Connectivity of Small Spheres}

\begin{figure}
\includegraphics[width=3.4in]{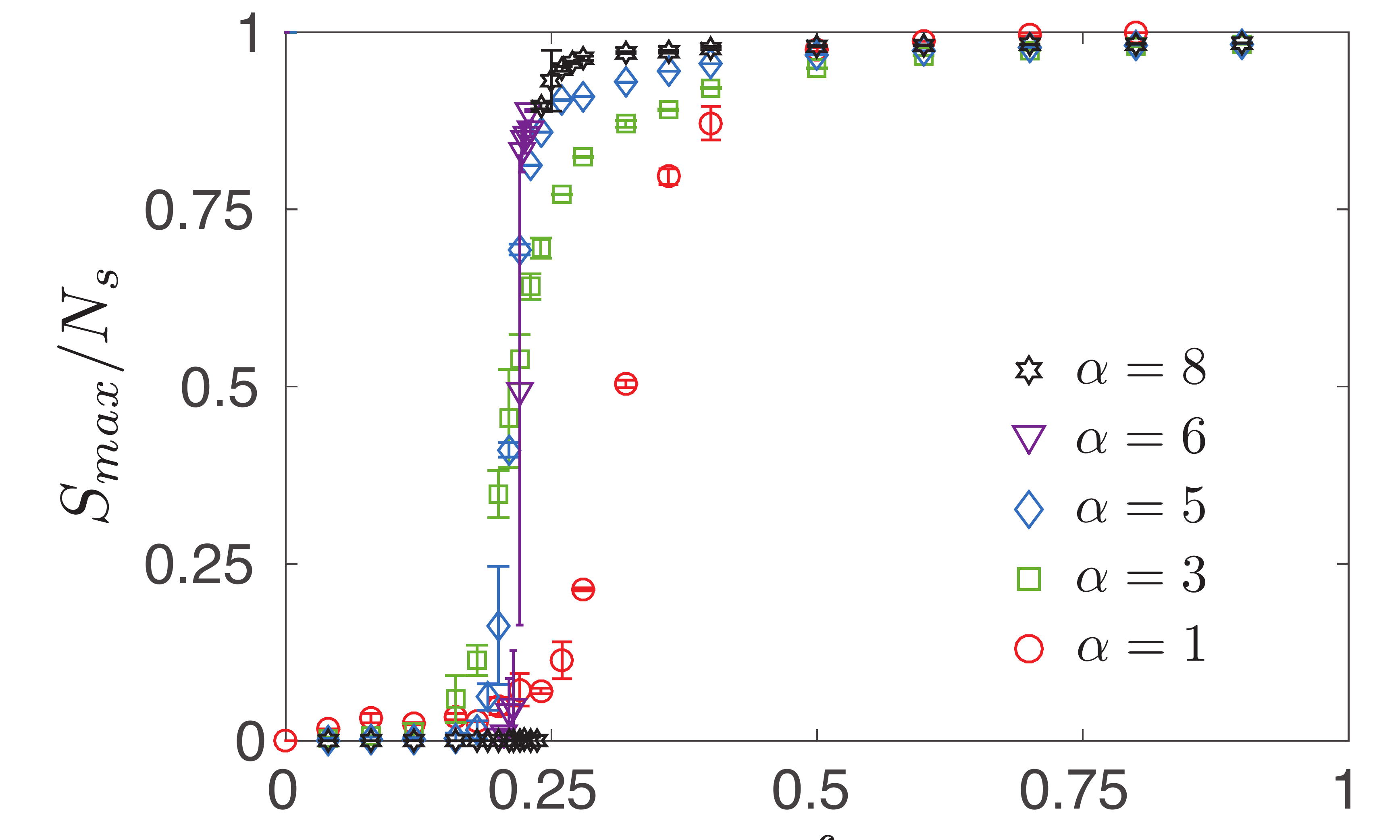}
\caption{\label{fig:7} The fraction of small spheres in the largest connected component of jammed small sphere contacts, plotted as a function of $f$ for various size ratios. }
\end{figure}

The small sphere sub-jamming transition outlined above has important implications for long-range structural correlations in binary sphere mixtures.  In particular, it suggests that global connectivity of the network of small-sphere contacts is critically sensitive to size ratio.  In Figure  \ref{fig:7}, we analyze this in terms of the fraction, $S_{max}/N_s$, of small spheres contained in the largest connected cluster of small sphere contacts, which includes the $S_{max}$ spheres.  For $\alpha \leq \alpha_c$ we find that  $S_{max}/N_s$ increases smoothly from 0 to 1 over a range of small sphere composition $f \approx 0.1 -0.4$.  We note that in the monodisperse limit, as $\alpha \to 1$, this approaches the problem of site percolation on a RCP contact network of monodisperse spheres, a problem that has received considerable previous study~\cite{Stauffer1994,Powell1979, Ahmadzadeh1982, Oger1986,Ziff2017}, motivated by conductivity studies in random mixtures of metallic and insulating particles~\cite{ Fitzpatrick1974,Ottavi1978, Takeuchi2012}. These earlier studies show that at $\alpha=1$ the percolation transition is continuous (i.e. a second-order transition, smeared out by finite-sized corrections), occurring at a fraction of $f \simeq 0.311$ ``occupied" sphere sites~\cite{Powell1979}.  Our results show that as $\alpha$ decreases towards 1 the maximal slope of $S_{max}/N_s$ occurs around $f\approx 0.32$ which is in close agreement to the percolation threshold for conducting versus non-conducting spheres of same size. The maximum slope would tend to infinity as number of spheres in the system reached the thermodynamic limit. 

We find that as $\alpha$ approaches $\alpha_c\simeq 5.75$ from below, the transition to $S_{max}/N_s \to 1$ sharpens considerably.  For $\alpha > \alpha_c$, the maximal slope of $S_{max}/N_s$ vs. $f$ apparently diverges (within resolution in $f$ increments), suggestive of a singular or discontinuous dependence on small sphere composition across the sub-jamming transition.  For example, for $\alpha=3$, $S_{max}/N_s$ rises from 0.1 to 0.9 over a composition increment $f=0.18-0.35$,  for $\alpha=5$ and $\alpha=6$ this same jump in $S_{max}/N_s$ occurs over $f=0.195-0.26$ and  $f=0.216-0.23$, respectively.  While the statistics of the small-sphere connectivity in random mixtures of binary spheres do not map strictly onto a percolation transition, or at least not within the most standard formulations of site percolation, the structural model suggested by Sec. \ref{sec: cooperativity} does suggest that a type of percolation of small-sphere connectivity occurs at large $\alpha$ along the sub-jamming transition line.  Here,``sites" are the interstitial simplices, whose enclosed small spheres are either jammed or unjammed at a given probability that increases with $f$ (Fig.  \ref{fig:5}F), and percolation occurs when neighboring jammed interstices form a connected cluster that spans the system.  While it remains unclear how to connect the two percolation scenarios (site percolation in the monodisperse RCP network at $\alpha =1$ and percolation of jammed large-sphere interstices for $\alpha \geq \alpha_c$) and further, what is the nature of percolation model (e.g. correlated percolation) the sub-jamming of large-sphere interstices, it is clear from Fig.~\ref{fig:7} that sub-jamming transition appearing for $\alpha \geq \alpha_c$ leads to a vastly enhanced sharpness (relative to the nearly monodisperse case) of the transition from disconnected to globally-connected small spheres along $f_{sub}(\alpha)$.

\section{\label{sec:Concl} Conclusion}

In summary, we give evidence of two distinct classes of binary sphere mixtures separated by a critical size ratio $\alpha_c$.  For nearly size symmetric spheres ($\alpha < \alpha_c$), local and global properties evolve smoothly with small sphere composition $f$.  Above the critical size asymmetry ($\alpha \geq \alpha_c$), we find evidence of a sharp transition from small-sphere poor to small-sphere rich packings, marked by an abrupt change in the fraction of jammed small spheres largely occupying the interstitial space between jammed large spheres along the line $f_{sub}(\alpha)$.     This demonstrates that the singular evolution of binary sphere mixtures suggested by the heuristic model of Furnas at  $\alpha \to \infty$ extends down to a finite range of size asymmetry.  We argue that the critical value of $\alpha_c\sim 5.75$ is related to the geometric criteria from small spheres to extend force contacts between large sphere interstices.  This distinction between large and small size asymmetry mixtures have important consequences for the design and engineering of materials composed by binary sphere mixtures, in particular, in conducting material composites where material properties are sensitive to global connectivity of like sphere (or unlike spheres)~\cite{Fitzpatrick1974, Ottavi1978, Bertei2011, Labastide2011, Renna2015}

At present, it remains to be understood, if there is an emergence of a divergent length scale (and scaling exponents) that characterize the ``critical end point" like behavior at $\alpha_c$.  Additional, questions remain about whether aspects of the apparent ``decoupling" of small- and large-size elements apply to random mixtures of other particle shapes.  For example, simulations of random rod-sphere mixtures at high aspect ratios  \cite{Kyrylyuk2010,Kyrylyuk2011} show signatures of similar structural transitions with variable composition of fine and large-scale elements. Better understanding of the universality of this transition could lead to widespread applications by improving structural and transport properties of granular materials.

\begin{acknowledgments}
The authors are grateful to J. Machta for helpful discussions.    This work was supported as part of the Polymer-Based Materials for Harvesting Solar Energy, an Energy Frontier Research Center funded by the U.S. Department of Energy, Office of Science, Basic Energy Sciences under Award DE-SC0001087.  Numerical simulations were performed on the UMass Shared Cluster at the Massachusetts Green High Performance Computing Center.
\end{acknowledgments}


 \bibliography{BinarySpheres}

\end{document}